\documentclass[a4paper,UKenglish]{lipics}

\usepackage{amsfonts}
\usepackage{color}
\usepackage{url}
\usepackage{ifthen}
\usepackage{graphicx}
\usepackage{caption}

\newcommand{\ignore}[1]{}

   {\begin{center}\begin{tabular}{l@@{\quad}l@@{\quad\quad\quad}l}}%
   {\end{tabular}\end{center}}%

\date{}

\title{A pilot study of the use of LogEx, lessons learned}

\author[1]{Josje Lodder}
\author[1]{Bastiaan Heeren}
\author[1,2]{Johan Jeuring}
\affil[1]{Faculty of Management, Science and Technology, Open University of the Netherlands,
    P.O.Box 2960, 6401 DL Heerlen, The Netherlands \\
  \texttt{Josje.Lodder@@ou.nl, Bastiaan.Heeren@@ou.nl}}
\affil[2]{Department of Information and Computing Sciences, \\ Universiteit Utrecht, The Netherlands \\
  \texttt{J.T.Jeuring@@uu.nl}}
\authorrunning{J. Lodder, B. Heeren, and J. Jeuring}

\Copyright{Josje Lodder, Bastiaan Heeren, and Johan Jeuring}


\serieslogo{logo_ttl}
\volumeinfo
  {M. Antonia {Huertas}, Jo\~ao {Marcos}, Mar\'ia {Manzano}, Sophie {Pinchinat}, \\
  Fran\c{c}ois {Schwarzentruber}}
  {5}
  {4th International Conference on Tools for Teaching Logic}
  {1}
  {1}
  {93}
\EventShortName{TTL2015}

\begin{document}

\maketitle
   
\begin{abstract}
LogEx is a learning environment that supports students in rewriting
propositional logical formulae, using standard equivalences. 
We organized a pilot study to prepare a large scale evaluation of the learning
environment. In this paper we describe this
study, together with the outcomes, which teach us valuable lessons for the large scale
evaluation. 
\end{abstract}

\section{Introduction}

Students learning propositional logic practice by solving different kinds of
exercises. Many of these exercises are solved stepwise. To support a student
solving such exercises an intelligent tutoring system can be very
effective~\cite{vanlehn}. At the Open University of the Netherlands we are developing a
learning environment (LE) LogEx\footnote{\url{http://ideas.cs.uu.nl/logex/}}, which
supports students in rewriting propositional logical formulae, using standard
equivalences. We intend to evaluate this LE with a large group of students
later this year, and to prepare this evaluation we organized a small scale
evaluation in December 2014~\cite{shute1993}. Detailed loggings of the learning environment
offer the possibility to analyze the way students use our LE. For example, in
an earlier study~\cite{Roijers} loggings of students working on normalizing
propositional logic expressions were used to construct a probability model of
the correctness of the use of rules. In the large scale study we want to perform,
our main focus is the question whether or not a student learns by using our
LE. We will compare different versions of LogEx that have more (or fewer)
feedback services. In the pilot study our main questions were: (1) do students learn by using the LE, and (2) what lessons can we learn
for a large scale evaluation.

This paper is organized as follows. In the next two sections we describe the
LogEx learning environment, and the experiment we performed with it. Section 4
summarizes the results of the assessment tests and loggings. We conclude with
lessons learned from the experiment.

\section{The LogEx learning environment}
\label{sec:LogEx}

\begin{figure}
\center\includegraphics[width=.85\textwidth]{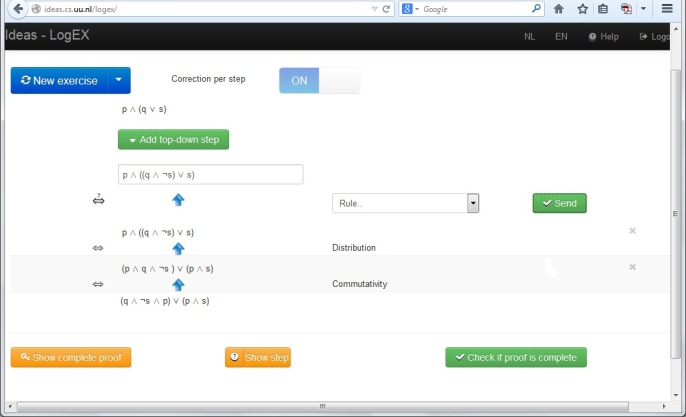}
\caption{Screenshot of LogEx.}
\label{screenshot}
\end{figure}

LogEx is a learning environment (LE) in which a student practices rewriting
propositional logical formulae, using standard equivalences. The LE contains
three kinds of exercises: rewriting a formula in DNF, in CNF, and proving the
equivalence of two formulae. A student enters her solution stepwise. In
exercises on equivalence proofs she has to motivate each step with a rule name;
in exercises on rewriting to normal form this motivation is optional. When
proving two formulae equivalent, a student can work both forwards and backwards.
Figure~\ref{screenshot} shows an example of a partial proof consisting of two
backward steps. A student can change the direction in which she is working at
any moment.

In the version we used in the pilot, correction per step is turned on. This implies
that a student receives feedback after each step. Feedback concerns syntax
errors, such as missing parentheses, or rule feedback. After a student has
entered a formula, the LE will try to recognize the rule that was used. If the LE can
detect a rule, it will compare this rule with the rule name provided by the
student, and give an error message if the wrong rule name was given. If no rule
is detected, the LE checks semantically whether or not the new and old formulae
are equivalent. If they are not equivalent, the LE uses a set of common
mistakes, also called buggy rules, to try to give informative feedback. For
example, if a student rewrites |not (p \/ q) \/ (not2 p /\ not q) \/ not q|
into |(not p \/ not q) \/ (not2 p /\ not q) \/ not q|, then the LE reports that
this step is incorrect, and mentions that when applying DeMorgan's rule, a
disjunction is transformed into a conjunction.

A student can ask for a hint (e.g.~perform a backward step or apply DeMorgan), a
next step, or a complete worked out solution, at any moment. The LE contains
solution strategies to calculate this feed forward. A student can choose between
exercises of different difficulty levels, or enter her own exercises. Feedback
and feed forward are available for all exercises. LogEx integrates improved
versions of earlier tools to rewrite formulas in disjunctive normal form
~\cite{lodder,LodderUsingIdeas} and to prove
equivalences~\cite{provingequivalences}. The main learning goals for which we want to use the LE are: After practicing with the LE a student
\begin{itemize} 
\item can recognize applicable rules 
\item can apply rules correctly 
\item can rewrite a formula in normal form 
\item can prove the equivalence of two formulae using standard equivalences
\item can demonstrate strategic insight in how to rewrite a formula in normal
form or prove an equivalence in an efficient way.
\end{itemize}

\section{The experiment}
\label{sec:experiment}
In the last decade we performed a number of small scale experiments with earlier
versions of our LE. Participants in these experiments were students of the Open
University of the Netherlands (OUNL). Although we learned quite a lot from these experiments,
they had two limitations. First, since OUNL is a distance
university, students evaluated the LE at home. This meant that we could not
observe students using the LE, and there was little control on the way how
students worked with the LE. Second, the students of OUNL are
rather heterogeneous. 

We performed a new experiment with our LE with a group of first-year students of
Utrecht University. This experiment had two goals:
\begin{itemize}
\item evaluate the use of the LE: do students achieve the learning goals as mentioned in Section~\ref{sec:LogEx} and is the support offered by the LE sufficient to reach these goals.
\item prepare for a large scale evaluation of the LE this year: is the
information about the LE sufficient to work with it, do students get enough time
to practice with the LE, do students get enough time to answer the pre-test
and post-test and is the logging adequate.
\end{itemize}
Before the experiment started, we organized a short introduction to the purpose
of the experiment and we explained the main features of the LE. Then students
got ten minutes to make a pre-test consisting of three exercises: prove that a
formula is a tautology, rewrite a formula into normal form, and prove that two
formulae are equivalent.
They practiced 75 minutes with the LE, and then made a
post-test comparable to the pre-test. We used a special version of the LE with a
fixed set of exercises: five on rewriting a formula in DNF, five in CNF, and
five on proving an equivalence. We logged all interactions of the students with
the LE. During the pre-test and post-test, students could make use of a paper sheet 
with the list of standard equivalences that were allowed to solve the problems.

Five students participated in the experiment, all male, age 18--22, from the
disciplines computer science, information science, and game technology. All
students took part in the course Logic in Computer Science, and had already
worked on the subjects covered in the LE in this course, except for proving an
equivalence using standard equivalences, which is not part of the learning goals
of the course.

\section{Results}
\begin{figure}[p]
\center\includegraphics[width=.8\textwidth]{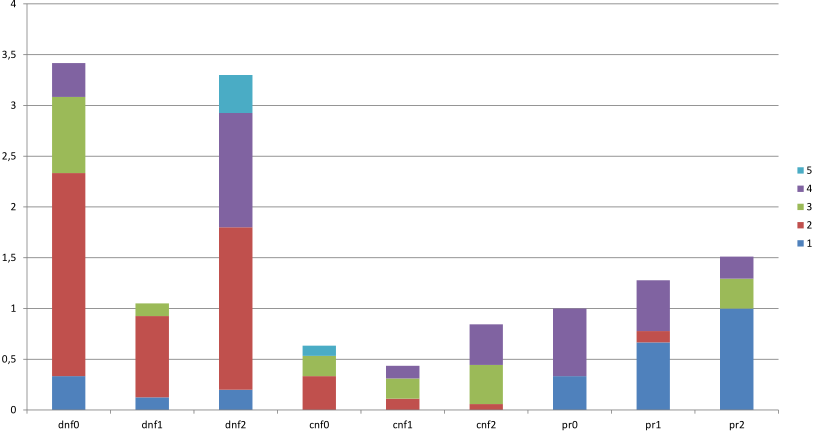}
\caption{Fraction of errors}
\label{errors}
\end{figure}

We analyzed the loggings of our LE. Only the first three normal form and proof
exercises were completed by almost all students, and we only include these
exercises in our results. Figure~\ref{errors} shows the number of erroneous
steps as a fraction of the total number of correct steps performed by the student. The x-axis displays the type and number of the exercise. All but one student completed the exercises in the order presented on this axis. Student 1 completed the exercises on proofs before
the exercises on CNF. Moreover, in the first three exercises on CNF this student
used only the next step button. He did complete the fourth CNF exercise
without help of the LE, but we did not include this
exercise in the figures. The figure shows that the students 2, 3, and 5 gradually 
make less mistakes, and this holds also for the last three exercises of student 4. 
The relatively high number of mistakes in the exercises dnf2 and pr2 can be explained by the 
difficulty of these exercises. Dnf2 asked for more complicated applications of the 
distribution rule. Pr2 had a rather simple solution, but students who did not see this solution got rather long formulae.

\begin{figure}[p]
\center\includegraphics[width=.8\textwidth]{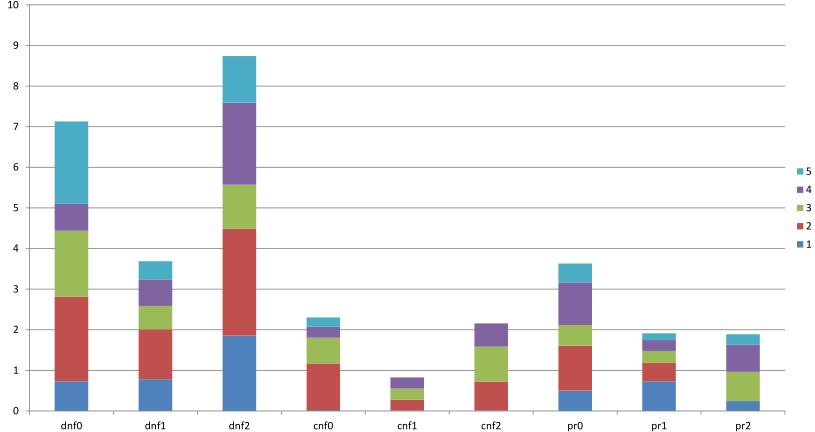}
\caption{Time per step (in minutes)}
\label{tijd}
\end{figure}

We also logged the time that a student needed to complete an exercise. In
Figure~\ref{tijd} we present the time per correct step in minutes. 
At the start students still learn how to use the LE, and this
likely partially accounts for the decrease of time needed to take a step. Most
students solved the post-test faster than the pre-test. This indicates that
students learn to solve the exercises faster by using the LE. 
The extra amount of time needed to solve pr0 can be explained by the fact that 
this type of exercise was new for this group of students.

\begin{figure}[p]
\center\includegraphics[width=.8\textwidth]{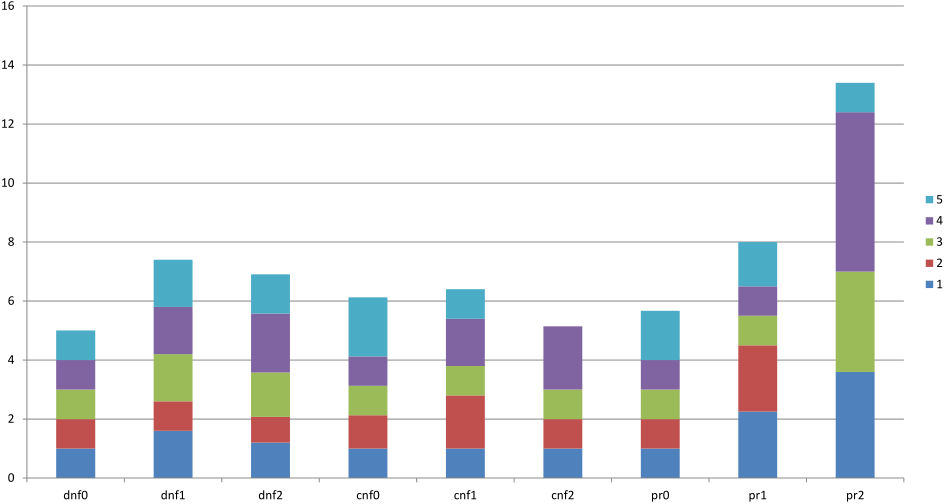}
\caption{Efficiency measured by the number of performed steps as a fraction of the number of steps in a worked out solution} 
\label{steps} 
\end{figure}

To evaluate the last learning goal, the development of strategic insight, we
compared the number of correct steps of the student solution with the number of
steps of the example solution generated by the LE. Here we took also into account how far
a solution was simplified. For example, in the first exercise, some students used
three steps to reach the normal form |(q /\ not r) \/ q \/ r|, others used 
four steps to reach the simplification |q \/ r|. The worked out solution also takes three steps to reach |(q /\ not r) \/ q \/ r| and four steps to reach |q \/ r|. Hence, in Figure~\ref{steps} we score 
both solutions as 1, namely $3/3$ resp. $4/4$. 
The outcome suggests that students do not learn to
solve the exercises more efficiently. We will discuss this outcome later. 

\begin{figure}[t]
\center
\includegraphics{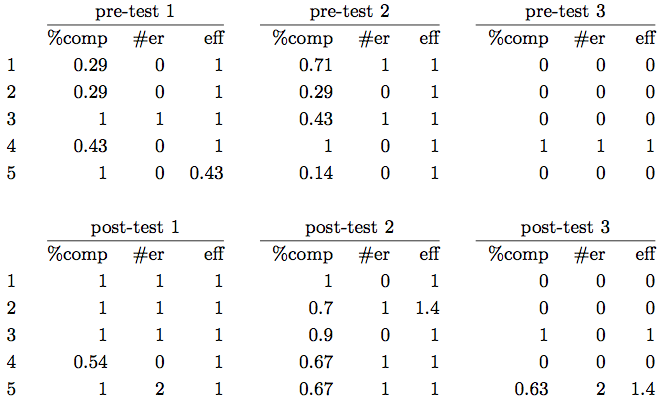}
\caption {Results of pre-test and post-test}{
  
\footnotesize
\begin{tabular} {l  p{12cm}}
\%comp      & number completed steps divided by total number of steps \\
\#er & number of errors\\
eff    & efficiency: number of steps performed by the student divided by number of steps in example solution \\
\end{tabular}
    }
\label{testresults}
\end{figure}

The results of the pre-test and post-test were not very informative. The main
positive result is that students completed more exercises in the post-test, see
Figure~\ref{testresults}. They made a few more mistakes in the post-test, but
this might be related to the number of steps they performed. Also, there was no
gain in efficiency. Because of the low number of students we could not compare
the difficulty of the pre-test and post-test. We think that the post-test was
slightly more difficult than the pre-test. The large scale evaluation will be
designed in such a way that we can compensate for different difficulties in 
pre-test and post-test.

\section{Lessons learned}

In this section we discuss the consequences of our analyses of the loggings and
the tests, together with observations we made during the evaluation session. 
\subsection{Do students reach the learning goals?}

The first question we posed in Section~\ref{sec:experiment} was: do students reach the learning goals and does LogEx sufficiently supports them to reach these goals.

\paragraph*{Recognizing applicable rules}
Students learn to recognize applicable rules, with two exceptions. LogEx admits
generalizations of DeMorgan and distribution rules. For example, it is allowed to
rewrite |not(p /\ q /\ r)| in |not p \/ not q \/ not r| in one step. These
generalizations were mentioned in the introduction to the evaluation, but they were not explicitly
present in the list of rules. Students did not use these generalized rules. A
second rule that was hardly used is absorption. This rule is not needed to
rewrite a formula in normal form, but it can simplify the calculations. Only one
of the students used this rule by himself, three others only after a hint
suggested to use absorption, and one student did not use this rule at all. We
have to think of a possibility to make students aware of the usefulness of this
rule. A possible solution might be that in case absorption is applicable,
but a student chooses another rule, LogEx will point out the possibility to
simplify the formula using absorption.

\paragraph*{Apply rules correctly}
LogEx does provide feedback at the rule level, and we find that this feedback
helps to achieve the second learning goal. In general the error
messages are sufficient for a student to correct mistakes. However, this is not
always the case. In case a student accidentally rewrites a formula into an
equivalent formula while making a mistake, no error specific message is given.
During the session we were asked several times by a student why his rewriting
was incorrect. Finally, in the loggings we found some examples where students
could not repair their mistakes directly in such a situation. In a next version
we will also provide error specific feedback when the new formula is equivalent
to the previous one.

Analysis of the loggings revealed some missing buggy rules, for example
rewriting of |(p \/ q) /\ (not p \/ not q)| into |F|.
From the loggings we also learned that error messages for syntax errors were not
always helpful: students sometimes need several attempts to correct a syntax
error.

\paragraph*{Rewrite a formula in normal form and prove the equivalence of two formulae}

The loggings and tests indicate that students do learn to rewrite a formula in
normal form. Students were able to complete the exercises without too much use of
the help button, and most students could finish the exercises on normal forms in
the post-test. Since time in the post-test was too short to complete all the
exercises, we can only use the loggings to draw conclusions about
proving equivalence. The loggings indicate that students also learn to solve this kind
of exercises.

\paragraph*{Demonstrate strategic insight}  

The loggings and tests do not show improvement on the last learning goal. 
A reason might be that students had to answer different kinds of exercises, which 
needed partially different strategies. A careful analysis of the loggings shows 
that this is not the only reason.  For example, one of the students
developed a personal strategy of introducing double negations combined with the
use of DeMorgan. In most cases this strategy was not effective, but since he got
no feedback on the use of this strategy, he kept using it, also in the post-test.
We think that there are at least two reasons why a student does not learn to
work more efficiently. The first reason is that LogEx does not provide feedback
on the strategic level, and hence gives no information about a strategy for
solving an exercise. This information is given implicitly by hints or next
steps, but only one student made use of hints or next steps. The possibility to
compare a solution with the complete solution was only used by one student.
Help avoidance is one of the known problems with
LEs~\cite{Vaessen,Vanlehn:2006,aleven03}. This might be a second reason that the
last goal was not met. Although in general students learn more when they have to
ask for help themselves~\cite{Vanlehn:2006}, in this case it seems necessary
that the system provides help without being asked. LogEx recognizes when a
student solution diverges from one of the possible paths of the strategy that we
implemented. In a next version we might provide a warning in such a case.
Alternatively, LogEx might warn a student if a solution is getting longer than
the worked-out solution. A third possibility is to postpone this warning until a
student has finished an exercise, but this might cause frustration.

\paragraph*{Other remarks concerning the use of the tool}

 To prevent unreadable
formulae and endless derivations, the use of associativity is implicit in LogEx. 
This means, for instance,  that a student does not have to introduce or change parentheses 
before an application of idempotency in a formula such as |q \/ p \/ p \/ s|. 
As a consequence, LogEx will consider |p \/ q \/ r| and |(p \/ q) \/
r| to be the same formula. There is no separate rule available to delete parentheses.  In the second DNF exercise most students reached the
normal form |q \/ (not p \/ q) \/ p|. At this point, the students tried to get
rid of the parentheses, but LogEx did not accept this. In a next version we will
have to introduce the possibility to delete parentheses.

Some other minor points we learned about the LE concern user friendliness.
Overall, students had no problems with the use of LogEx. However, we observed a
student copying and pasting a previous formula when he wanted to correct the
formula he was editing. He had not noticed that the mini-keyboard in the user 
interface contains an undo button.

\subsection{What lessons can we learn for the large scale evaluation?}

The use of a pilot study is an important principle in the design of evaluation
studies~\cite{shute1993}. Overall, the evaluation went well, but students need
more time for the pre-test and post-test. We log all requests and messages
between the user and the domain reasoner, but some actions are not yet logged at
this moment. For example, LogEx offers the possibility to undo some steps in a
proof, but the use of the undo button is not logged. We can only indirectly
assume that a student removed part of her proof from the fact that the old
formula in a rewriting is not equal to the new formula. Without knowing if and
where students use the undo button, it is very hard to draw conclusions about
the effectiveness of the student solutions.

To draw conclusions about the learning of the students during the use of the tool, it is necessary that the order in which students make the exercises is fixed.

The instruction about the use of commutativity was not clear. LogEx admits
commutative variants of the standard equivalences. For example, the rewriting of
|phi /\ (psi \/ chi)| in |(phi /\ psi)\/ (phi /\ chi)| is in the
list of standard equivalences, and LogEx also allows the variant
where |(psi \/ chi)/\ phi| is rewritten in |(psi /\ phi)\/ (chi /\ phi)|.
However, LogEx considers the rewriting of |(psi \/ chi)/\ phi| in |(phi /\
psi)\/ (phi /\ chi)| to be a combination of distributivity and commutativity,
which cannot be performed in one step. Students did perform these kind of steps
without realizing why LogEx did not accept the step.

\section{Conclusion}
The pilot indicates that with some adaptations, especially in feedback on the
strategic level, LogEx can be a helpful LE for students who practice rewriting
logical formulae. The large scale evaluation later this year will have to
confirm these findings. The pilot was useful for the design of the large scale
evaluation, in particular with respect to the timing of the components, the
instruction, and the loggings.

\bibliographystyle{plain}
\bibliography{strategies,FeedbackServices}

\end{document}